\documentclass[fleqn,usenatbib]{mnras}
\usepackage{newtxtext,newtxmath}
\usepackage[T1]{fontenc}
\usepackage{ae,aecompl}
\usepackage{graphicx}	
\usepackage{amsmath}	
\usepackage{amssymb}	
\usepackage{soul}
\usepackage{newtxtext}
\usepackage[english]{babel}
\usepackage{booktabs}
\usepackage{graphicx,color}
\usepackage{epstopdf}

\newcommand\reallywidehat[1]{\arraycolsep=0pt\relax
\newcommand{\stkout}[1]{\ifmmode\text{\sout{\ensuremath{#1}}}\else\sout{#1}\fi}

\begin{array}{c}
\stretchto{
  \scaleto{
    \scalerel*[\widthof{\ensuremath{#1}}]{\kern-.5pt\bigwedge\kern-.5pt}
    {\rule[-\textheight/2]{1ex}{\textheight}} 
  }{\textheight} %
}{0.5ex}\\           
#1\\                 
\rule{-1ex}{0ex}
\end{array}
}
\title[Twin Peaks in CoDECS]{Weak Lensing Peaks in Simulated Light-Cones:
  Investigating the Coupling between Dark Matter and Dark Energy}
\author[Giocoli C. et al. 2017]{\parbox{\textwidth}{
    Carlo Giocoli$^{1,2,3}$\thanks{E-mail:\href{mailto:carlo.giocoli@unibo.it}
      {carlo.giocoli@unibo.it}}, Lauro Moscardini$^{1,2,3}$, Marco Baldi$^{1,2,3}$, 
      Massimo Meneghetti$^{2,1,3}$, 
      Robert B. Metcalf$^{1,2}$ 
  }\\ \\
$^1$Dipartimento di Fisica e Astronomia, Alma Mater Studiorum Universit\`{a} di 
Bologna, via Gobetti 93/2, 40129 Bologna, Italy\\ 
$^2$INAF-OAS  Osservatorio di Astrofisica e Scienza dello Spazio di Bologna, via Gobetti 93/3, 40129, Bologna, Italy \\ 
$^3$INFN - Sezione di Bologna, viale Berti Pichat 6/2, 40127, Bologna, Italy 
}
\pubyear{2017}

\begin{document}
\label{firstpage}
\pagerange{\pageref{firstpage}--\pageref{lastpage}}
\maketitle

\begin{abstract}

  In this paper, we study the statistical properties of weak lensing
  peaks in light-cones generated from cosmological simulations. In
  order to assess the prospects of such observable as a cosmological
  probe, we consider simulations that include interacting Dark Energy
  (hereafter DE) models with coupling term between DE and Dark
  Matter. Cosmological models that produce a larger population of
  massive clusters have more numerous high signal-to-noise peaks;
  among models with comparable numbers of clusters those with more
  concentrated haloes produce more peaks. The most extreme model under
  investigation shows a difference in peak counts of about $20\%$ with
  respect to the reference $\mathrm{\Lambda CDM}$ model. We find that
  peak statistics can be used to distinguish a coupling DE model from
  a reference one with the same power spectrum normalisation. The
  differences in the expansion history and the growth rate of
  structure formation are reflected in their halo counts, non-linear
  scale features and, through them, in the properties of the lensing
  peaks. For a source redshift distribution consistent with the
  expectations of future space-based wide field surveys, we find that
  typically seventy percent of the cluster population contributes to
  weak-lensing peaks with signal-to-noise ratios larger than two, and
  that the fraction of clusters in peaks approaches one-hundred
  percent for haloes with redshift $z\leq 0.5$. Our analysis
  demonstrates that peak statistics are an important tool for
  disentangling DE models by accurately tracing the structure
  formation processes as a function of the cosmic time.

\end{abstract}


\section{Introduction}
In the standard cosmological model, most of the energy in the
Universe, approximately $70\%$, is in an unknown form, termed Dark
Energy (hereafter DE) which has a negative pressure.  This component
is responsible for the late time accelerated expansion as measured by
many observations
\citep{pelmutter99,riess99,riess04,riess07,schrabback10,betoule14}.
About $25\%$ of the energy content is in a different unknown component
termed Dark Matter (DM), whose presence has been mainly inferred from
its gravitational effects given that it seems not to emit nor absorb
detectable levels of radiation
\citep{zwicky37,rubin80,bosma81a,bosma81b,rubin85}.

Following the standard scenario, cosmic structures form as a
consequence of gravitational instability.  Dark matter overdensities
contract and build up into so-called dark matter haloes
\citep{white78,white79}.  Small systems collapse first when the
universe is denser and then merge together to form more massive
objects \citep{tormen98a,lacey93,lacey94}.  Galaxy clusters sit at the
top of this hierarchy as the latest nonlinear structures to form in
our Universe
\citep{kauffmann93,springel01b,springel05b,wechsler02,vandenbosch02,wechsler06,giocoli07}.

The large amount of dark matter present in virialized systems and
within the filamentary structure of our Universe is able to bend the
light emitted by background objects \citep{bartelmann01}. Because of
this, the intrinsic shapes of background galaxies appear to us weakly
distorted by gravitational lensing.  Since lensing is sensitive to the
total mass of objects and independent of how the mass is divided into
the light and dark components of galaxies, groups and clusters, it
represents a direct and clean tool for probing the distribution and
evolution of structures in the Universe.

When light bundles emitted from background objects travel through high
density regions like the centres of galaxies and clusters, the
gravitational lensing effect is strong (SL): background images appear
strongly distorted into gravitational arcs or divided into multiple
images \citep{postman12,hoekstra13,meneghetti13,limousin16}.  On the
other hand, when light bundles transit the periphery of galaxies or
clusters, background images are only slightly distorted and the
gravitational lensing effect is termed weak (WL)
\citep{amara12,radovich15}.  In this way weak gravitational lensing
represents an important tool for studying the matter density
distributed within large scale structures.  A large range of source
redshifts allows one to tomographically probe the dark energy
evolution through the cosmic growth rate as a function of redshift
\citep{kitching14,kohlinger15} \citep[for a review
  see][]{kilbinger14}.  Great efforts and impressive results have been
reached by weak lensing collaborations like CFHTLens
\citep{fu08,benjamin13} and KiDS \citep{hildebrandt17}.  Some tensions
may still exist between these measurements and the ones coming from
the Cosmic Microwave Background \citep{planckxxiv}.  Hopefully, wide
field surveys from space will help to fill the gap between low- and
high-redshift cosmological studies and shed more light onto the dark
components of our Universe.

Gravitational lensing will be the primary cosmological probe in
several experiments that will start in the near future, like LSST
\citep{lsst} and the ESA space mission
Euclid\footnote{\href{https://www.euclid-ec.org}{https://www.euclid-ec.org}}
\citep{euclidredbook}. Recently, the Kilo Degree Survey (KiDS)
collaboration presented a series of papers devoted to the shear peak
analysis of $\sim 450$ deg$^2$ of data \citep{hildebrandt17}.  They
emphasised that peak statistics are a complementary probe to cosmic
shear analysis which may break the degeneracy between the matter
density parameter, $\Omega_m$, and $\sigma_8$, the power spectrum
amplitude expressed in term of the root-mean-square of the linear
density fluctuation smoothed on a scale of $8$ Mpc$/h$.  In
particular, \citet{shan17} analyzed the convergence maps reconstructed
from shear catalogues using the non-linear \citet{kaiser93} inversion
\citep{seitz95}.  They showed that, given their source redshift
distribution, peaks with signal-to-noise larger than three are mainly
due to systems with masses larger than $10^{14}M_{\odot}/h$. However,
the source distribution in the KiDS observations corresponds to a
galaxy number density of only $7.5$ gal. per square arcmin at a median
redshift of $z=0.6$.  This low number density of galaxies prevented
them from performing a tomographic analysis. Within the same
collaboration, by using reconstructed maps from simulations,
\citet{martinet17} confirmed the importance of combining peak and
cosmic shear analyses.  In particular they pointed out that
cosmological constraints in the $\Omega_m$-$\sigma_8$ plane coming
from low signal-to-noise peaks are tighter than those coming from the
high-significance ones.

The strength of peak statistics in disentangling cosmological models
has been discussed in the last years by several authors. In particular
\citet{maturi11} have inspected the effect of primordial
non-Gaussianity, which impacts the chance of projected large scale
structures varying the peak counts. \citet{pires12} demonstrated that
peak counts are the best statistic to break the $\sigma_8$-$\Omega_m$
degeneracy among the second-order weak lensing statistics.
\citet{reischke16} have suggested that the extreme value statistic of
peak counts can tighten even more the constraints on cosmological
parameters.

In this work we will study weak lensing peak statistics in a sample of
non-standard cosmological models which are characterised by a coupling
term between dark energy and dark matter.  We will discuss the
complementarity of peak statistics with respect to cosmic shear and
examine the information on non-linear scales from high significance
peaks.  We will discuss also the importance of tomographic analysis of
peak statistics as tracers of the growth and the expansion history of
the universe.

The paper is organised as follows: in Section~\ref{secmethodsns} we
present the numerical simulations analysed and introduce how weak
lensing peaks have been identified. Statistical properties of peaks
are reviewed in Section~\ref{secwlpp}, while the connection between
galaxy clusters and peaks is discussed in Section~\ref{secpclusters}.
We conclude and summarise in Section~\ref{summary}.

\section{Methods and Numerical Simulations}
\label{secmethodsns}

\subsection{Numerical Simulations of Dark Energy Models}

In this work we use the numerical simulation dataset presented by
\citep{baldi12b} and partially publicly available at this url:
\href{http://www.marcobaldi.it/web/CoDECS_summary.html}{http://www.marcobaldi.it/web/CoDECS\_summary.html}.
The simulations have been run with a version of the widely used N-body
code {\small GADGET} \citep[][]{springel05a} modified by
\citet{baldi10a}, which self-consistently includes all the effects
associated with the interaction between a DE scalar field and
$\mathrm{CDM}$ particles.  The $\mathrm{CoDECS}$ suite includes
several different possible combinations of the DE field potential --
the exponential \citep[][]{lucchin85,wetterich88} or the
$\mathrm{SUGRA}$ \citep[][]{brax99} potentials for example -- and of
the coupling function which can be either constant or exponential in
the scalar field \citep[see e.g.][]{baldi11b}. For more details on the
models we refer to \citet{baldi12b}.

In particular we use some simulations of the $\mathrm{L-CoDECS}$
sample ($\mathrm{\Lambda CDM}$, $\mathrm{EXP003}$, $\mathrm{EXP008e3}$
and $\mathrm{SUGRA003}$) plus $\mathrm{\Lambda CDM-HS8}$ that is a
$\mathrm{\Lambda CDM}$ simulation with the same cosmological
parameters as $\mathrm{\Lambda CDM}$ but with a value of $\sigma_8$
equal to the one of $\mathrm{EXP003}$.  The $\mathrm{\Lambda CDM-HS8}$
simulation has been run in order to study how the effect of the
coupling between DE and DM can be disentangled from a pure
Cosmological Constant model with the same power spectrum
normalisation.  A summary of the considered simulations with their
individual model parameters is given in Table~\ref{tab:models}
\begin{table*}
\caption{The list of the cosmological models considered in the present
  work and their specific parameters.  All the models have the same
  amplitude of scalar perturbations at $z_{\rm CMB}\approx 1100$, but
  have different values of $\sigma _{8}$ at $z=0$.  In short, $\alpha$
  is a parameter describing the slope of the scalar field potential,
  $\beta(\phi)$ is the coupling function describing the rate of
  energy-momentum exchange with dark matter, and $w_{\phi}(z=0)$ is
  the effective equation of state parameter ($p/\rho$). See
  \citet{baldi12b} for details.}
\label{tab:models}
\begin{tabular}{llcccc}
\hline
Model & Potential  &  
$\alpha $ &
$\beta (\phi )$ &
$w_{\phi }(z=0)$ &
$\sigma _{8}(z=0)$\\
\hline
$\mathrm{\Lambda CDM}$ & $V(\phi ) = A$ & -- & -- & $-1.0$ & $0.809$ \\
$\mathrm{\Lambda CDM-HS8}$ & $V(\phi ) = A$ & -- & -- & $-1.0$ & $0.967$ \\
$\mathrm{EXP003}$ & $V(\phi ) = Ae^{-\alpha \phi }$  & $0.08$ & $0.15$ & $-0.992$ & $0.967$\\
$\mathrm{EXP008e3}$ & $V(\phi ) = Ae^{-\alpha \phi }$  & $0.08$ & $0.4 \exp [3\phi ]$& $-0.982$ & $0.895$ \\
$\mathrm{SUGRA003}$ & $V(\phi ) = A\phi ^{-\alpha }e^{\phi ^{2}/2}$  & $2.15$ & $-0.15$ & $-0.901$ & $0.806$ \\
\hline
\end{tabular}
\end{table*}

We also use the information about the halo catalogue computed for each
simulation snapshot using a Fried-of-Friend (FoF) algorithm with
linking parameter $b=0.2$ times the mean inter-particle separation. At
each simulation snapshot, within each FoF group we also identify
gravitationally bound substructures using the \textsc{subfind}
algorithm \citep{springel01b}.  \textsc{subfind} searches for
overdense regions within a FoF group using a local SPH density
estimate, identifying substructure candidates as regions bounded by an
isodensity surface that crosses a saddle point of the density field,
and testing that these possible substructures are physically bound
with an iterative unbinding procedure.  For both FoF and
\textsc{subfind} catalogues we select and store systems with more than
20 particles, and define their centres as the position of the particle
with the minimum gravitational potential.  It is worth noting that
while the subhaloes have a well-defined mass that is the sum of the
mass of all particles belonging to them, different mass definitions
are associated with the FoF groups. We define as $M_{\rm FoF}$ the sum
of the masses of all particles belonging to the FoF group and as
$M_{200}$ the mass around the FoF centre enclosing a density that is
$200$ times the critical density of the universe at the corresponding
redshift.

\begin{figure}
  \includegraphics[width=\hsize]{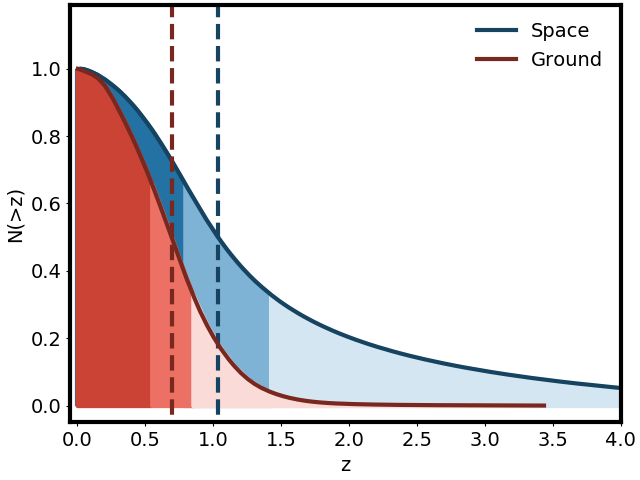}
  \caption{Cumulative normalised probability functions for two source
    redshift distributions, termed Space and Ground.  While the Ground
    distribution has been build to match the source redshift
    distribution of CFHTLens \citep{kilbinger13}, the Space one
    corresponds to the distribution adopted by
    \citet{boldrin12,boldrin16} as expected from a wide field survey
    from space, like Euclid.  The two dotted lines mark the median
    source redshifts in the two cases, while the different coloured
    regions below the curves indicate the redshift ranges in which one
    third of the galaxies are expected.\label{nzfig}}
\end{figure}

To compare the expected results for surveys from ground and space, we
adopt in our analyses two different distribution functions of sources,
shown in Fig.~\ref{nzfig}.  The red (blue) curve, normalised to unity,
mimics the probability distribution of sources as (expected to be)
observed from ground (space) photometric survey. In particular the red
curve corresponds to the redshift distribution from CFHTLens
\citep{kilbinger13}, while the blue curve corresponds to the
distribution adopted by \citet{boldrin12,boldrin16}.  The latter has
been obtained using a simulated observation with the \textsc{SkyLens}
code \citep{meneghetti08,bellagamba12,rasia12} and identifying with
\textsc{SExtractor} \citep{bertin96} sources $3$ times above the
background rms.  The two dashed vertical lines, red and blue, mark the
median redshift from ground and space, respectively.  The regions
shaded in three gradations of colour enclose the redshift ranges where
we have one-third of the number density of sources for the two
corresponding distributions.  As can be seen, the source distribution
from space moves toward higher redshifts with a considerable tail that
extends beyond $z=2$: the expectations from space-based observations
suggest a gain of at least a factor of two in the number of galaxies
per square arcmin with measurable shapes.  We reasonably assume a
total number density of $18$ and $33$ galaxies per arcmin$^2$ for a
ground and space experiment, respectively.

\begin{figure*}
  \includegraphics[width=0.95\hsize]{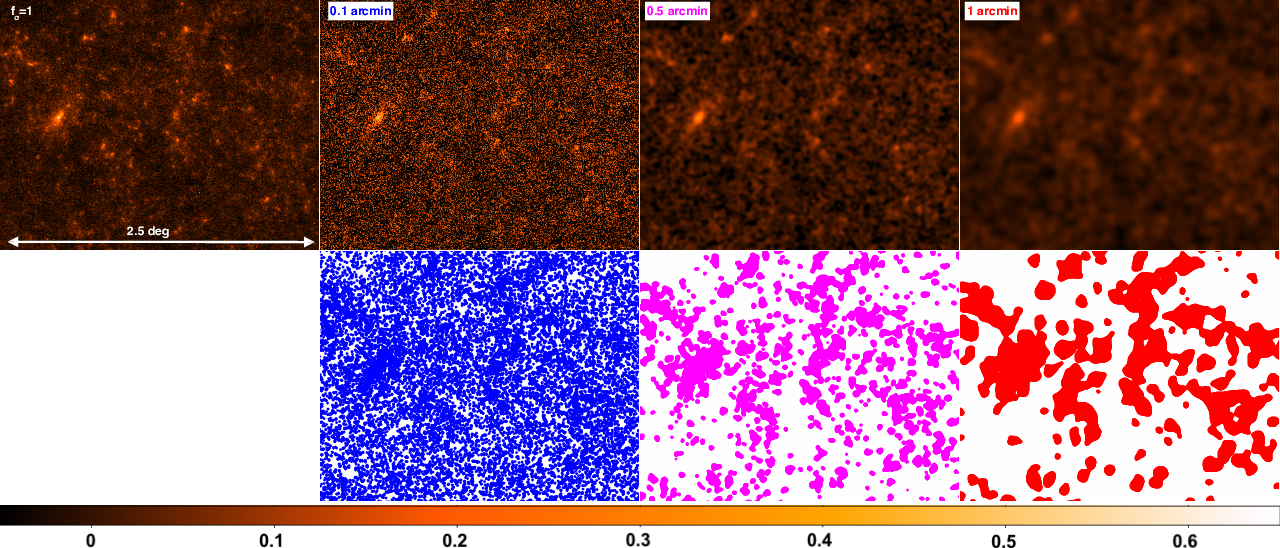}
  \caption{Noised and smoothed convergence maps considering different
    choices for the filter $\theta_F$. The top left panel shows the
    original convergence map.  The other top panels, moving from left
    to right, display the convergence maps artificially noised and
    filtered considering $\theta_F$ $0.1$, $0.5$ and $1$ arcmin. The
    bottom sub-panels display the regions, in the corresponding maps,
    above the noise level. \label{figthetaF}}
\end{figure*}

\subsection{Light-Cone Reconstruction and Peak Detection}

We perform our weak lensing peak detection using convergence maps for
different source redshifts and for various cosmological models.  We
employ the \textsc{MapSim} routine developed by \citet{giocoli15} to
construct $25$ independent light-cones from the snapshots of our
numerical simulations.  We build the lens planes from the snapshots
while randomising the particle positions by changing sign of the
comoving coordinate system or arbitrarily selecting one of the nine
faces of the simulation box to be located along the line-of-sight.  If
the light-cone reaches the border of a simulation box before it has
reached a redshift range where the next snapshot will be used, the box
is re-randomised and the light-cone extended through it again.  The
lensing planes are built by mapping the particle positions to the
nearest pre-determined plane, maintaining angular positions, and then
pixelizing the surface density using the triangular-shaped cloud
method.  The selected size of the field of view is $5\times 5$
sq. degrees and the maps are resolved with $2048 \times 2048$ pixels,
which corresponds to a pixel resolution of about 8.8 arcsec.  Through
the lens planes we produce the corresponding convergence maps for the
desired source redshifts using the \textsc{glamer} code
\citep{metcalf14,petkova14,giocoli16a}.

As done by \citet{harnois-deraps15b}, for the lens planes stacked into
the light-cones we define the \emph{natural} source redshifts as those
lying at the end of each constructed lens planes.  By construction our
light-cone has the shape of a pyramid where the observer is located at
the vertex and the base extends up to the maximum redshift chosen to
be $z=4$.

In wide field weak lensing analysis it is worth mentioning that
intrinsic alignments (IAs) of galaxies may bias the weak lensing
signal.  However \citet{shan17} have shown that considering an
Intrinsic Alignment (IA hereafter) amplitude as computed from the
cosmic shear constraints by \citet{hildebrandt17}, the relative
contribution of IA to the noise variance of the convergence is very
small and well bellow $0.6\%$ with respect to randomly oriented
intrinsic ellipticities.  Thus, to first approximation, we assume that
the galaxies are intrinsically randomly oriented.

Noise can affect cosmological lensing measurements and results in
possible biased constraints on cosmological parameters. One of the
methods used to suppress the noise in reconstructed weak lensing
fields is smoothing.  Since weak gravitational lensing is by
definition a weak effect, it is necessary to average over a sufficient
number of source galaxies in order to obtain a measurement.  Because
of the central limit theorem, after smoothing the statistical
properties of the noise field are expected to be close to a Gaussian
distribution.  For the noise and the characterisation of the
convergence maps we follow the works of \citet{lin15a,lin15b}.  The
convergence maps $\kappa(x,y)$ that we produce from our ray-tracing
procedure are only characterised by the discreteness of the density
field sampled with collisionless particles: the so-called particle
noise.  However, to mimic the presence of galaxy shape noise, from
which the convergence map is inferred from real observational data, we
add to $\kappa(x,y)$ a noise field $n(x,y)$ that accounts for this. If
we assume that the intrinsic ellipticities of the source galaxies are
uncorrelated we can describe $n(x,y)$ as a Gaussian random field with
variance:
\begin{equation}
\sigma_{\rm noise}^2 = \dfrac{\sigma^2_{\epsilon}}{2} \dfrac{1}{2 \pi
  \theta^2_F n_g}\;,
\end{equation}
where $\sigma_{\epsilon}=0.25$ is the rms of the intrinsic ellipticity
of the sources, $n_g$ the galaxy number density and $\theta_F$
represents the smoothing scale of a Gaussian window function filter,
that we apply to the noised convergence map to suppress the pixel
noise \citep{lin15a,matilla16,shan17}. We indicate with $k_{n,F}$ the
noised and filtered convergence map.  Consistent with the choice made
by other authors we adopt a scale of $1$ arcmin for the smoothing
scale which represents the optimal size to isolate the contribution of
massive haloes typically hosting galaxy clusters. For descriptive
purposes, in the top left panel of Figure~\ref{figthetaF} we display
the convergence map with an aperture of $2.5$ deg on a side and
$z_s=1.12$. In the three panels on the left the noise has been added
and the map has been smoothed assuming different choices of
$\theta_F$, $0.1$, $0.5$ and $1$ arcminutes, from left to right,
respectively. The coloured regions in the bottom panels mark the
pixels in the image above that are above the noise level
$f_{\sigma}=1$ with:
\begin{equation}
f_{\sigma} = \dfrac{\kappa_{n,F}}{\sigma_{\rm noise}}\,.
\end{equation}
From the figure we can see that peaks identified in the convergence
fields with small values of $\theta_F$ are dominated by false
detections caused by the noise level. For larger $\theta_F$ values the
peak locations consistently follow the locations of the interposed
halos within the field-of-view.

\begin{figure*}
  \includegraphics[width=0.8\hsize]{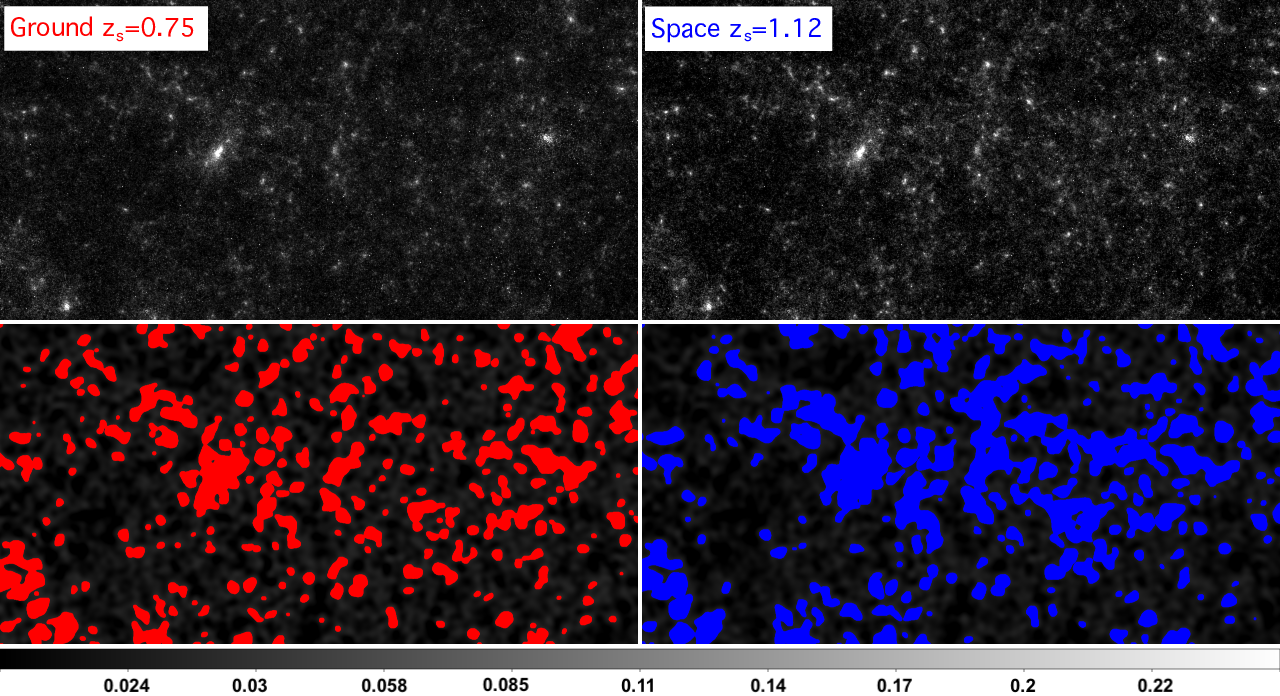}
  \caption{\label{figkappaSG}Top panels: convergence maps for one
    light-cone realisation of the $\mathrm{\Lambda CDM}$ simulation
    assuming sources at two fixed redshifts -- corresponding to the
    median redshifts for the space- and ground-based redshift
    distributions here considered. Bottom panels: pixels above the
    corresponding noise level $\sigma_{\rm noise}$. The scale of the
    field-of-view along the x-axis of the displayed regions is
    approximately $3$ degrees large.}
\end{figure*}

In our analysis we consider two natural source redshifts corresponding
to $z_s=0.75$ and $z_s=1.12$ that are the medians of the two source
redshift distributions as displayed in Fig.~\ref{nzfig} \footnote{We
  remind the reader that our distributions are supposed to mimic, in
  an optimistic way, a space- and ground-based experiment; in addition
  we point out that the source redshift distribution for the Euclid
  ESA Mission \citep{kitching16} is expected to have a median redshift
  of galaxies for shape measurement $z_m=0.9$.}.  The top panels of
Fig.~\ref{figkappaSG} displays the convergence maps of a light-cone
realisation from the $\mathrm{\Lambda}$CDM simulation considering
these two source redshifts. The bottom panels shows the pixels in the
corresponding maps $\kappa_{n,F}(x,y)$, noised and smoothed with
$\theta_F=1$ arcmin to account for observational effects with
$f_{\sigma}\geq1$
\footnote{Contrary to many peak studies we choose to indicate the peak
  height above the noise with $f_{\sigma}$ instead of $\nu$ since the
  latter is typically used in some of our previous works for
  $\delta^2_c(z)/\sigma^2(M)$.}
  
We characterise the peak properties for a given threshold $f_{\sigma}$
as following: ($i$) we identify all the pixels above $f_{\sigma}$
times the noise level, ($ii$) we join them to the same peak group
using a two-dimensional friend-of-friend approach adopting the pixel
scale as linking length parameter, ($iii$) we define the coordinate of
the peak centre according to the location of the pixel with the
maximum value and the area as related to the number of pixels that
belong to the group times the pixel area; we term our peak
identification algorithm
\textsc{TwinPeaks}\footnote{\href{https://www.youtube.com/watch?v=V0cSTS2cTmw}{https://www.youtube.com/watch?v=V0cSTS2cTmw.}}:
while for small values of the signal-to-noise threshold $f_{\sigma}$
some peaks are twins, for large values of $f_{\sigma}$ they become
distinct and isolate. We want to emphasise that, as discussed, the
peak identification method depends on the resolution of the
convergence map -- constructed from simulations or reconstructed using
the shear catalogue of an observed field of view. Being interested in
displaying and discuss relative differences in the counts and in the
properties of the peaks for various Dark Energy models, all the maps
have been created to have the same pixel resolution: field-of-view of
$5$ deg by side are resolved with $2048 \times 2048$ pixels,
consistently noised and smoothed using the same parameter choices.

\begin{figure*}
  \includegraphics[width=0.8\hsize]{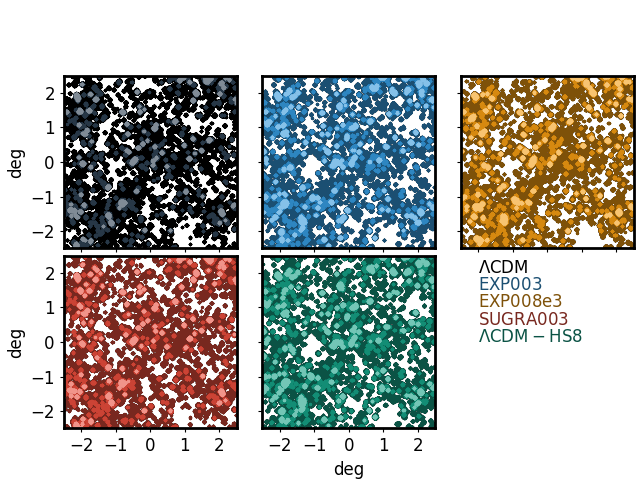}
  \caption{\label{figpeaksCoDECS}Examples of the weak lensing peak
    locations in the different cosmological models, from the map
    computed assuming a space source redshift distribution with
    $z_s=1.12$. The maps include shape noise and are smoothed with a
    Gaussian filter with scale $\theta_F=1$ arcmin.  The different
    coloured panels show various cosmological models as indicated in
    the label.  Within each panel the three gradations of colour mark
    the regions in the field-of-view which are $1$, $3$ and $5$ times
    above the expected noise level $\sigma_{\rm noise}$.}
\end{figure*}

\section{Weak Lensing Peak Properties in Coupled DM-DE Models}
\label{secwlpp}
We run the complete and self-consistent \textsc{TwinPeaks} pipeline on
all light-cones generated for the various cosmological models:
$\mathrm{\Lambda CDM}$, $\mathrm{EXP003}$, $\mathrm{EXP008e3}$,
$\mathrm{SUGRA}$ and $\mathrm{\Lambda CDM-HS8}$.  In all cases we have
considered two fixed source redshifts $z_s=0.75$ and $z_s=1.12$ (that
are the median source redshifts of the two considered source redshift
distributions) with a number density of galaxies of $18$ and $33$ per
square arcmin for the ground- and space-based observations,
respectively.  As an example, in Figure~\ref{figpeaksCoDECS} we
display the \textsc{TwinPeaks} results for light-cones derived from
the same random realisation of initial conditions at $z=99$ for the
five different cosmological models, colour coded as displayed in the
figure legend: black, blue, orange, red and green refer to
$\mathrm{\Lambda CDM}$, $\mathrm{EXP003}$, $\mathrm{EXP008e3}$,
$\mathrm{SUGRA003}$ and $\mathrm{\Lambda CDM-HS8}$, respectively.  In
this case, we show the results for $z_s=1.12$; in each panel the three
gradations of colours mark the regions which are $1$, $3$ and $5$
times above the noise level, considering a filter size $\theta_F= 1$
arcmin.

\subsection{Peak Counts}
\begin{figure*}
  \includegraphics[width=0.45\hsize]{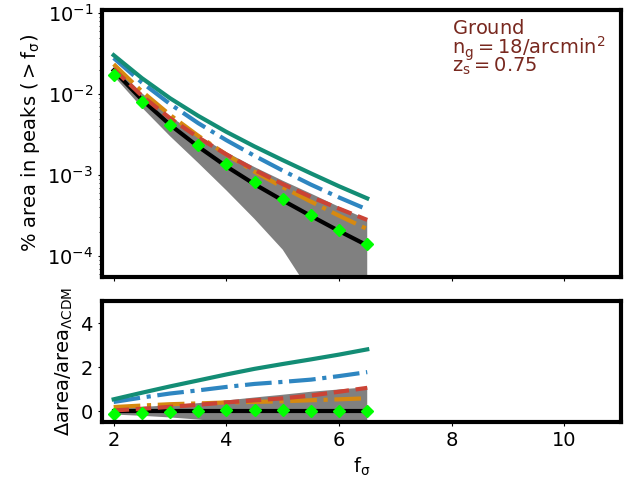}
  \includegraphics[width=0.45\hsize]{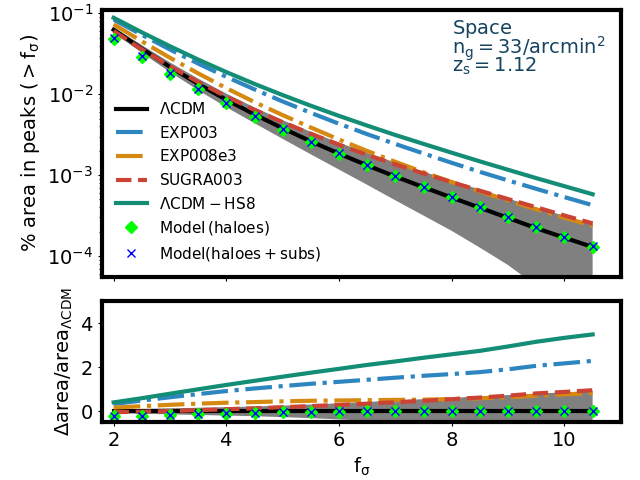}  
  \caption{\label{fitareanoiseCoDECS}Left and right panels show the
    fraction of the area covered by weak lensing peaks as a function
    of the threshold of the noise level, for the various cosmological
    models for a space and a ground based source redshift
    distribution, respectively. The grey area bracketing the black
    lines shows the variance of the different light-cone realisations
    for the $\mathrm{\Lambda CDM}$ model. The various coloured curves
    display the measurement done on the light-cones of the different
    cosmological models.  Green diamonds and blue crosses display the
    predictions obtained using our halo and halo plus subhalo models,
    discussed in the Appendix.}
\end{figure*}

Figure~\ref{fitareanoiseCoDECS} displays the fraction of the area
occupied by peaks as a function of the signal-to-noise level
$f_{\sigma}$, for the various cosmologies.  Each curve corresponds to
the average value computed on the $25$ different light-cone
realisations. Left and right panels display the results for a ground
and space analysis, respectively.  The outcomes for the various
cosmological models are shown using different colours.  The grey
region bracketing the measurements of the $\mathrm{\Lambda CDM}$ model
(black curve) shows the variance of the different light-cone
realisations. The variance for the other models is similar and then
not shown for clarity reasons.  The corresponding bottom panels
present the relative difference in the measured area in peaks with
respect to the reference $\mathrm{\Lambda CDM}$ model as a function of
the signal-to-noise value $f_{\sigma}$.  The green diamonds show the
predictions from our halo model formalism for the standard
$\mathrm{\Lambda CDM}$ model, described in more details in the
Appendix.  We notice that the model describes quite well the
predictions of the corresponding cosmological model, it captures
within few percents the behaviour for large values of the
signal-to-noise ratio. The blue crosses (present only on the right
panel) show the results of our model where we also include the
presence of subhaloes. As described by \citet{giocoli17} we treat them
as Singular Isothermal Spheres.  From a more detailed analysis we
highlight that subhaloes boosts the weak lensing peaks at most $3$
percent. This is due to two main reasons: ($i$) subhaloes are
typically embedded in more massive haloes whose contribution to the
convergence map is stronger and ($ii$) their presence may be washed
out by the noise and the smoothing of the convergence map.  From the
bottom panels we see that the higher peaks allow for a better
discrimination between different cosmological models, while for low
values of $f_{\sigma}$ the peaks trace mainly projected systems and
filaments.  At about $f_{\sigma}=6$ the two most extreme models EXP003
and $\mathrm{\Lambda CDM-HS8}$ show a positive difference of about
$15-20\%$ while at $f_{\sigma}=10$ - attainable for a space
observation with a large number density of background galaxies -- of
approximately $25-30\%$, in the regime where peaks are not dominated
by the shape noise. The fraction of area in peaks for the EXP008e3 and
SUGRA models are situated at almost $1$ $\sigma$ away from the
$\mathrm{\Lambda CDM}$ one.  It has also been pointed out by
\citet{maturi10} who showed that weak-lensing peak counts are
dominated by spurious detections up to signal-to-noise ratios of $3-5$
and that large scale structure noise can be suppressed using an
optimised filter.  For large $f_{\sigma}$ we detect the non-linear
scales (typically for angular modes with $l > 10^2$) where galaxy
clusters are located, making peak statistic complementary to cosmic
shear measurements \citep{shan17}.  We can also see that observations
from space should resolve peaks with a much higher resolution than
ground-based ones and also resolve peaks with much higher
signal-to-noise ratio where the difference between the various
cosmological models is largest.  Comparing the figure with the cosmic
shear forecast analyses on the same cosmological models by
\citet{giocoli15} we notice that high signal-to-noise peak statistics
is able to differentiate more the various dark energy models.  This
suggests that future wide field surveys like Euclid will be excellent
for this type of analyses, binding much more the cosmological models
not only in the $\Omega_m$-$\sigma_8$ planes but also in the dark
energy equation of state.

In Figure~\ref{fitcountsnoiseCoDECS} we display the number of peaks
above a given threshold of the signal-to-noise level; data points and
colours are the same as in Fig.~\ref{fitareanoiseCoDECS}.  From the
figure we notice that the trend of the peak counts is very similar to
that of the area in peaks as previously discussed.  The
$\mathrm{\Lambda CDM-HS8}$ is very distinct from the $\mathrm{\Lambda
  CDM}$ model in peak counts, showing also a different behaviour with
respect to the $\mathrm{EXP003}$ model which has the same power
spectrum normalisation. Peak statistics traces the different growth of
structures and expansion histories. From the bottom panels we can
notice that the $\mathrm{\Lambda CDM}$ model with high $\sigma_8$
predicts much more weak lensing peaks: this model has much more haloes
which are much more concentrated due to their higher formation
redshift.  In general a higher peak abundance in weak lensing fields
is mainly due to a combined effect of the projected halo mass function
in the light-cones and to the redshift evolution of the
mass-concentration relation.

\begin{figure*}
  \includegraphics[width=0.45\hsize]{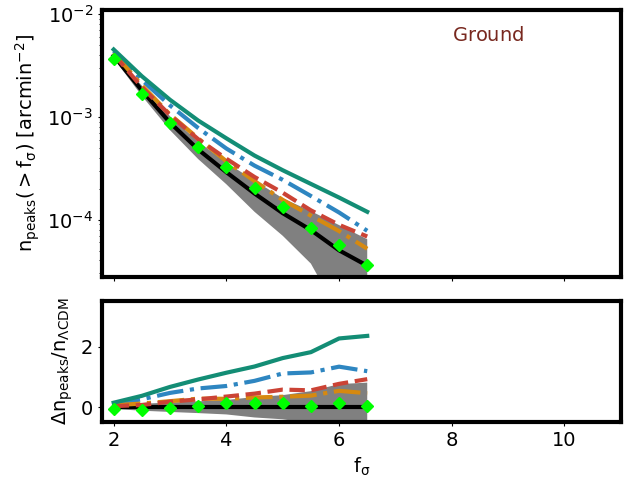}
  \includegraphics[width=0.45\hsize]{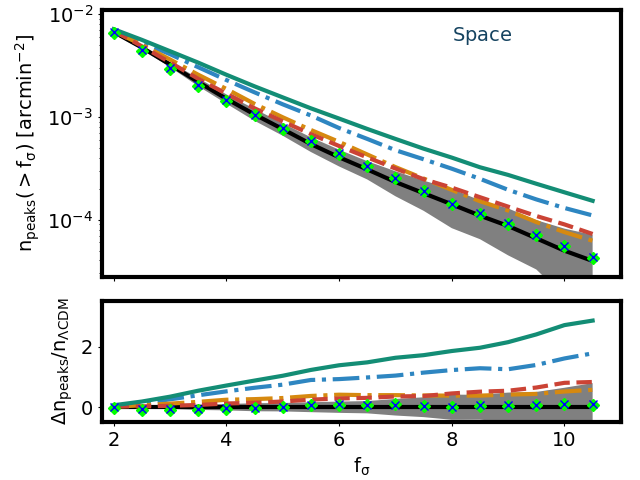}  
  \caption{Peak counts above a given noise $f_{\sigma}$ level for the
    various cosmological models.  Data points, lines and panels are as
    in Figure~\ref{fitareanoiseCoDECS}.\label{fitcountsnoiseCoDECS}}
\end{figure*}

\begin{figure*}
  \includegraphics[width=0.45\hsize]{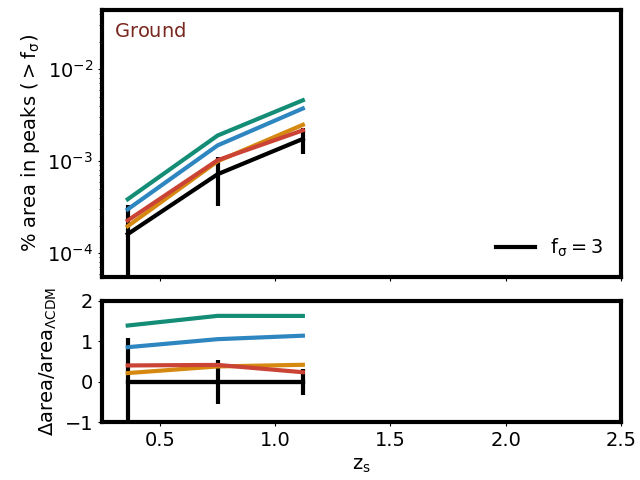}
  \includegraphics[width=0.45\hsize]{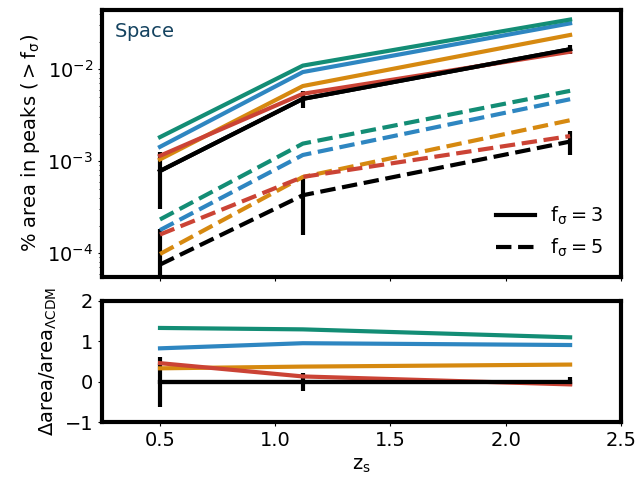}
  \caption{Fraction of the area in peaks as a function of the source
    redshift for space- (left) and ground-based (right) observations.
    Solid lines show the area above $3$ times the noise level, while
    the dashed ones consider peaks above $f_{\sigma}=5$. In particular
    for the ground-based experiment we display only the case for
    $f_{\sigma}=3$ since peaks with higher values of the noise are
    poorly resolved because of the number density of 8 galaxies per
    arcmin$^2$, per redshift bin. Various colours refer to the
    different cosmological models, as indicated in previous
    figures. \label{figtomography}}
\end{figure*}

Results presented until now considered sources located at a fixed
redshifts.  However weak lensing tomographic analyses provide the
possibility of tracing the structure formation process as a function
of redshift and can be an important constraint on the growth factor
and on the dark energy equation of state.  This can be possible as
long as we have a reasonable number of background galaxies per
redshift bin.  In order to perform a weak lensing peak analysis as a
function of redshift, both for the space- and ground-based cases we
divide the corresponding source redshift distribution in three
redshift bins that contain one-third of the total expected number
density of galaxies. As mentioned before those bins in redshift are
displayed with different colour gradations in Figure~\ref{nzfig}.  In
Figure \ref{figtomography} we present the fraction of the area in
peaks above a given threshold $f_{\sigma}=3$ as a function of the
source redshift for the various cosmological models and the two
experiments: from ground (left) and space (right): they have $6$ and
$11$ galaxies per arcmin$^2$ per bin, respectively.  For the space
case we also show the measurement for high peaks with $f_{\sigma}=5$
(dashed lines), that are not properly resolved for the ground based
experiment because of the low number density of background sources.
The black error bar corresponds to the rms in the measurements for the
reference $\mathrm{\Lambda CDM}$ model.  The tomographic peak analysis
illustrates the capability of following the structure formation
processes for the different cosmological models. While for the
ground-based case the maximum redshift considered is $z\approx 1.1$,
from space we can go up to $z\approx 2.3$. As in the previous
discussions both the EXP003 and $\mathrm{\Lambda CDM-HS8}$ present the
largest differences in peaks with respect to the reference
$\mathrm{\Lambda CDM}$ model.  For example the right panel displays
that the $\mathrm{SUGRA003}$ model has at high redshift an area in
peaks very similar to the $\mathrm{\Lambda CDM}$ cosmology, while at
low redshifts (as it can also be noticed in the left panel) the area
in peaks is larger than the corresponding one in the standard
model. This is actually consistent with the fact that
$\mathrm{SUGRA003}$ is a bouncing model characterised by a different
evolution of both the growth factor and the Hubble function
\citep[see][]{baldi12c}. Tomographic peak statistics will be a
powerful tool for discriminating dark energy models from standard
cosmological constant, being able to self-consistently trace the
growth of structures, and more specifically -- as we will discuss in
the next section -- of galaxy clusters as a function of the cosmic
time.

\section{Galaxy Clusters and Weak Lensing Peaks}
\label{secpclusters}

The results presented in the last section show that weak lensing peaks
tend to be located close to high-density regions of the projected
matter density distribution and that simulations based on the halo
model describe quite well both the peak area and number counts as a
function of the signal-to-noise ratio. The fact that the contribution
of subhaloes to the weak lensing peaks is negligible also suggests
that clusters, and line-of-sight projections of haloes, represent the
main contribution to high peaks in the convergence maps.

In this section we will discuss the correlation between peaks and
galaxy clusters present within the simulated light-cones, and try to
shed more light on the connection between high peaks and massive
haloes. We tag a halo as a contributor to a peak if its centre of mass
has a distance smaller then $1$ pixel from a peak above a certain
signal-to-noise value $f_{\sigma}$.

\begin{figure*}
 \includegraphics[width=0.32\hsize]{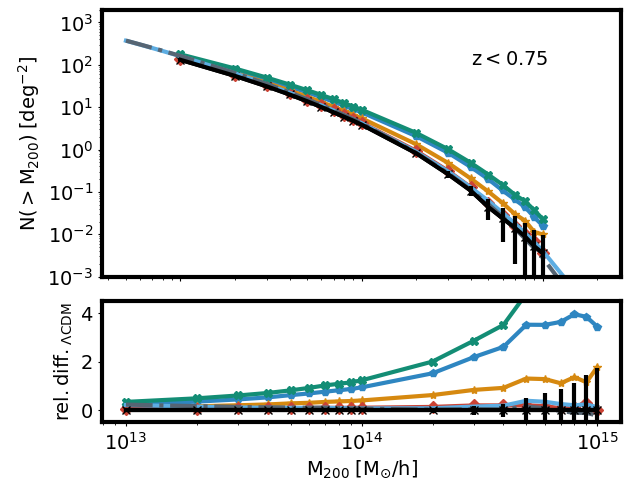}
\includegraphics[width=0.32\hsize]{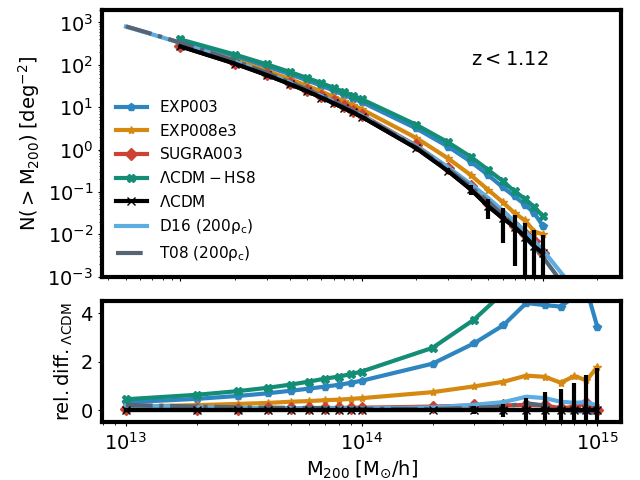}
 \includegraphics[width=0.32\hsize]{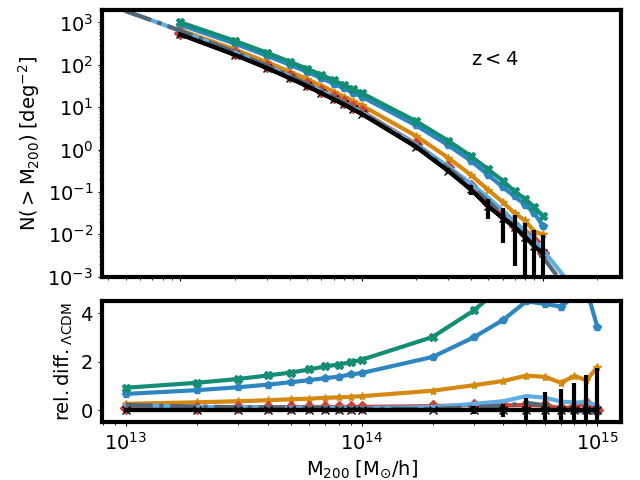}
\caption{Halo mass function per square degree within the simulated
  light-cones up to redshifts $z=0.75$, $z=1.12$ and $z=4$ from left
  to right respectively. The data points show the measurements in the
  various cosmological models -- with Poisson error bars displayed
  only for the $\mathrm{\Lambda CDM}$ model. The light-blue and
  dark-grey curves show the predictions from the \citet{despali16} and
  the \citet{tinker08} mass function for the $\mathrm{\Lambda CDM}$
  cosmology.\label{fighalomassfunction}}
\end{figure*}

In Figure~\ref{fighalomassfunction} we display the cumulative halo
mass function per square degree within the constructed light-cones,
for the various cosmological models, up to $z=0.75$, $z=1.12$ and
$z=4$ from left to right, respectively.  For the halo mass we use
$M_{200}$, the mass enclosing $200$ times the critical density of the
universe at the same redshift.  For comparison, in each panel the
light-blue and dark-grey curves display the predictions by
\citet{despali16} and \citet{tinker08} for the $M_{200}$ mass
definition. The bottom panels show the relative difference of the
counts with respect to the measurement in the standard
$\mathrm{\Lambda CDM}$ simulation. From these panels we can notice
that the integrated halo mass function of the $\mathrm{SUGRA003}$
model is very similar to the $\mathrm{\Lambda CDM}$ \citep[the
  $\mathrm{SUGRA003}$ model has been actually constructed to result in
  such similarity at low redshifts, see][for a detailed discussion on
  this issue]{baldi11b}. However the number of peaks in this model is
quite different (as shown in Fig.~\ref{fitareanoiseCoDECS}
and~\ref{fitcountsnoiseCoDECS}) and comparable to the peak counts in
EXP008e3.  This is a clear signature of the halo properties
\citep{cui12b,giocoli13}: clusters in the bouncing $\mathrm{SUGRA003}$
model form at higher redshifts and are typically very
concentrated. This translates in higher and more numerous peaks in the
convergence field. This is a confirmation that peak statistics is very
sensitive not only to the initial power spectrum but also to the
non-linear processes that characterise halo formation histories and
that may help disentangling models that would appear degenerate in
other observables as the halo mass function.  This is in agreement
with the finding obtained by \citet{shan17}: peak statistics gives
complementary constraints with respect to cosmic shear in the
$\sigma_8-\Omega_m$ plane, and in this case, as we have shown, also in
the extended parameter space of coupled Dark Energy cosmologies.

\begin{figure*}
\includegraphics[width=0.32\hsize]{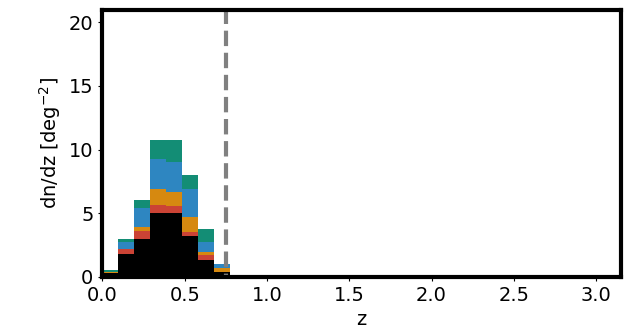}
\includegraphics[width=0.32\hsize]{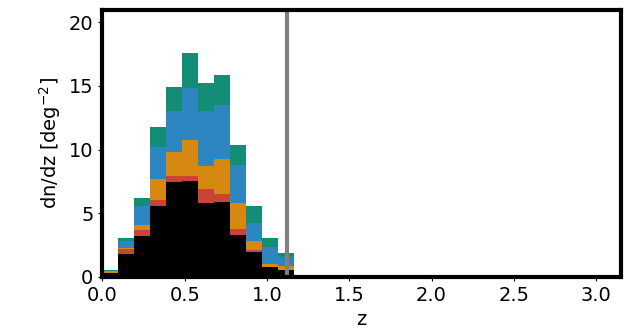}
\includegraphics[width=0.32\hsize]{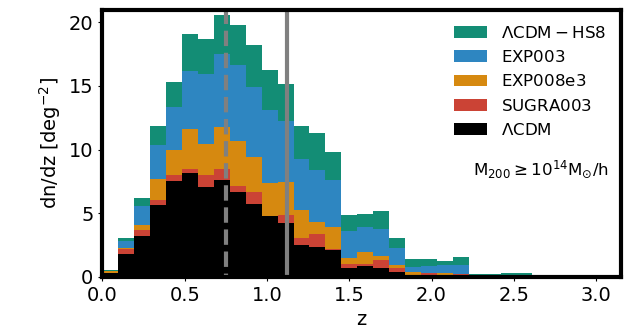}
\caption{Differential number density of clusters with $M_{200}\geq
  10^{14}M_{\odot}/h$ per unit of square degree for the various
  cosmological models. Left and central panels show the redshift
  distributions of clusters in peaks with $f_{\sigma}=2$ for the
  ground and space experiments, respectively. Right panel displays the
  redshift distribution of clusters for each cosmological model within
  our constructed light-cones. Solid and dashed vertical lines show
  the source redshift for the two considered cases ground and
  space-based: $z_s=0.75$ and $z_s=1.12$,
  respectively. \label{figclustersz}}
\end{figure*}

The three panels in Figure~\ref{figclustersz} show the redshift
distribution of clusters in the light-cones with mass
$M_{200}\geq10^{14}M_{\odot}/h$ as a function of redshift.  Left and
central panels display the redshift distribution of systems that fall
into peaks with $f_{\sigma}=2$ for the ground- and space-based
experiment, respectively; the right panel, instead, shows the
distribution of the whole cluster population within the constructed
past-light-cones.  Dashed and solid vertical lines mark $z_s=0.75$ and
$z_s=1.12$, respectively. In these figures it is possible to see that
the number of clusters in the $\mathrm{SUGRA003}$ model is quite
similar to $\mathrm{\Lambda CDM}$ one while large differences are
present in the counts with respect to the $\mathrm{\Lambda CDM}$ with
high $\sigma_8$ and $\mathrm{EXP003}$.

\begin{figure}
\includegraphics[width=\hsize]{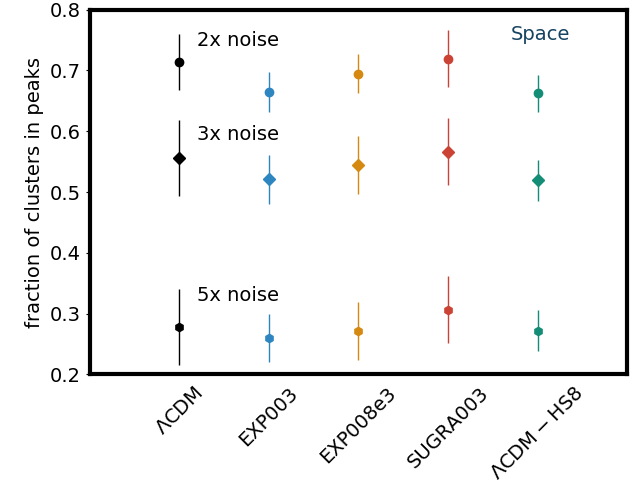}
\includegraphics[width=\hsize]{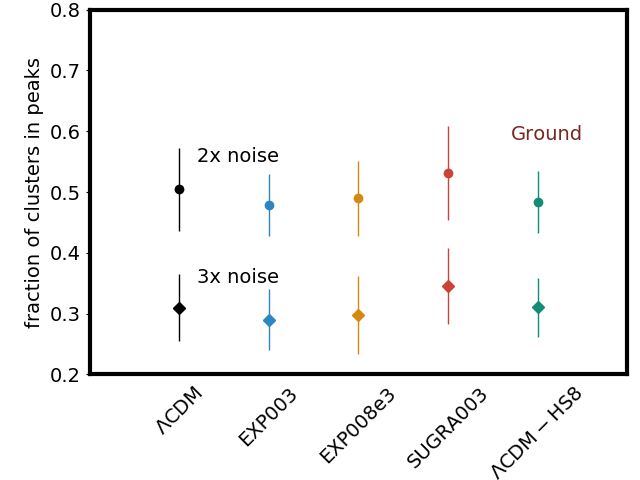}
\caption{Fraction of clusters in peaks for a space- and ground-based
  analysis of the weak lensing simulations (top and bottom panel
  respectively) for the various cosmological models considered in this
  work.  In the top panel we consider all the clusters with
  $M_{200}\geq 10^{14}M_{\odot}/h$ up to the source redshift of
  $z_s=1.12$, while in the bottom panel up to $z_s=0.75$.  Different
  data points display the fractions of those systems whose centre of
  mass falls within high convergence pixels which are part of weak
  lensing peaks above a given threshold value
  $f_{\sigma}$.\label{figfractionclpeaks1}}
\end{figure}

\begin{figure}
\includegraphics[width=\hsize]{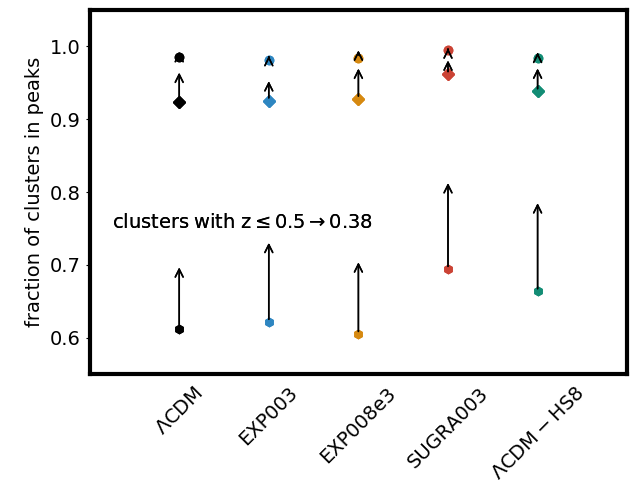}
\caption{Fraction of clusters for a space-based analysis with
  $M_{200}\geq 10^{14}M_{\odot}/h$ and $z\leq0.5$ in peaks with
  convergence values above $f_{\sigma}=2$, $3$ and $5$ from top to
  bottom data points, respectively.  The arrow on each data point
  shows the corresponding fraction of clusters in peaks when we select
  systems with $z\leq 0.38$ -- the peak of the lensing kernel for
  $z_s=1.12$.
\label{figfractionclpeaks05}}
\end{figure}

Top and bottom panels in Figure~\ref{figfractionclpeaks1} display the
fraction of clusters corresponding to weak lensing peaks for space-
and ground-based experiments, respectively. In both panels we show the
fraction of clusters in peaks above various weak lensing noise levels,
for the different cosmological models, colour coded as in the other
figures.  The considered source redshifts for the space- and
ground-based experiments are $z_s=0.75$ and $z_s=1.12$, respectively,
and that those also correspond to the maximum cluster redshift we
consider; moreover we consider clusters with masses above
$M_{200}\geq10^{14}M_{\odot}/h$. We notice that for the space
experiment we find that almost $55\%$ ($70\%$) of the clusters with
$z<1.12$ are in peaks $3$ ($2$) times above the noise level, while for
the ground-based experiment it is nearly $30\%$ ($50\%$) of all
clusters with $z<0.75$.  We remind the reader that for a cluster to be
within a peak it is necessary that its projected centre of mass falls
in a pixel of the corresponding map that is above the desired
threshold: by definition each peak, depending on its shape, may or not
contain more than a halo with $M_{200}\geq10^{14}M_{\odot}/h$.  The
halo contribution to the corresponding weak lensing field is weighted
by the lensing distance $D_{lens}\equiv D_{ls} D_l/D_s$ (where $D_l$
$D_s$ and $D_{ls}$ are the angular diameter distances observer-lens,
observer-source and source-lens, respectively) so that haloes, even if
they have the same mass, contribute differently to the lensing signal
depending on their redshift: for example, considering $z_s=1.12$ the
lensing distance $D_{lens}$ peaks around $z=0.38$.  This is more
evident in Figure~\ref{figfractionclpeaks05} where we show the
fraction of clusters with $z\leq 0.5$ in peaks above different
thresholds of the noise level, for the space case.  The fraction of
haloes with $M_{200}\geq10^{14}M_{\odot}/h$ and $z\leq 0.5$ in peaks
with $f_{\sigma}=2$ is close to unity.  The arrow on each data point
shows the corresponding fraction of clusters in peaks when we select
systems with $z\leq 0.38$ -- the peak of the lensing kernel for
$z_s=1.12$.

\begin{table*}
\caption{Number of clusters with $M_{200}\geq 10^{14}M_{\odot}/h$
  (with $z<1.12$, second and third columns, and $z\leq 0.5$, forth and
  fifth columns) in peaks above the threshold of $f_{\sigma}=2$ in the
  various cosmological models.  Second and fourth columns display the
  number of clusters in the various models up to redshift $1.12$ and
  $0.5$, respectively. On the other side, third and fifth columns
  present the corresponding cluster counts in peaks with
  signal-to-noise ratio $f_{\sigma}=2$, while the number between
  parentheses refers to the number of clusters in peaks for which the
  centre of mass corresponds with the pixel with the highest
  value. Numbers refer to the sum over $25$ different light-cone
  realisations, for a total of $625$ sq. deg., for each cosmology.
\label{tab1}}
\begin{tabular}{lllcrr}
$ $ & n. cl. $z<1.12$ $\rightarrow$& n. cl. in peaks  (with $\Delta \theta_{\rm cl, peak}=0$)  &||& n. cl. $z\leq 0.5$ $\rightarrow$& n. cl. in peaks (with $\Delta \theta_{\rm cl, peak}=0$)  \\ \hline \hline
$\mathrm{\Lambda CDM}$ & $3730$ & \textbf{2655 (90)}     &||&  1207  & \textbf{1188 (53)}  \\ 
$\mathrm{EXP003}$                              & 8223 & 5460 (158) &||&  2130 & 2088 (83)  \\
$\mathrm{EXP008e3}$                           & 5523 & 3834 (102) &||& 1602 & 1576 (57)  \\
$\mathrm{SUGRA003}$                          & 4069 & 2926 (130)  &||& 1314 & 1308 (77)  \\ 
$\mathrm{\Lambda CDM-HS8}$ & 9684 & 6410 (191) &||& 2429  & 2391 (110)  \\ \hline
Model  & \emph{3730} &  \textbf{2634 (124)} &||& \emph{1207} & 1201 (69) \\  \hline \hline
\end{tabular}
\end{table*}

\begin{figure}
\includegraphics[width=\hsize]{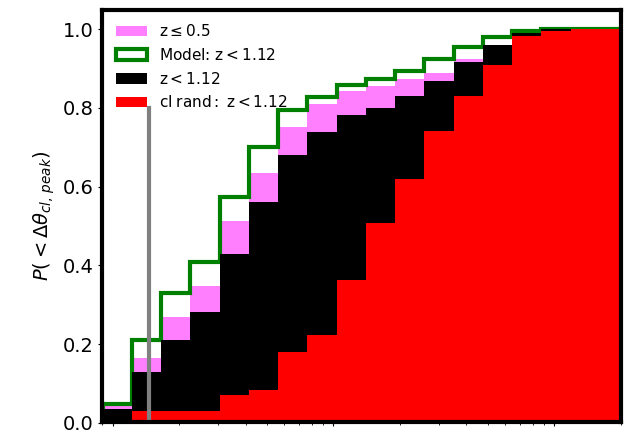}
\includegraphics[width=\hsize]{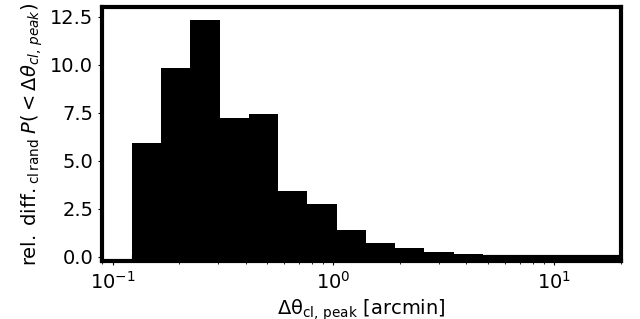}
\caption{Normalised cumulative distribution of the distance between
  the centre of mass of the clusters in peaks and the closest pixel
  with highest value. The results refer to the $\mathrm{\Lambda CDM}$
  cosmology.  This result does not have important cosmological
  dependence, all the other models possess a similar distribution. The
  vertical grey line indicates the angular scale corresponding to the
  pixel size in the convergence map. The green histogram refers to the
  measurements with respect the convergence maps constructed using our
  fast weak lensing halo model formalism, discussed in details in the
  Appendix.  The red histogram shows the cumulative distribution of
  the distance between peaks and clusters, when the latter are assumed
  to have random position within the field-of-view.  The bottom panel
  displays the relative difference between the black and the red
  histograms.\label{figdistpc}}
\end{figure}

The correlation between weak lensing peaks (above a given threshold)
and clusters represents a promising statistics to identify regions in
the plane of the sky where clusters are more likely to be found.  In
Figure~\ref{figdistpc} we display the normalised cumulative
distribution of the angular distances $\Delta \theta_{\rm cl, peak}$
between the cluster centre of mass and the location of the closest
pixel with the highest value of the convergence with $f_{\sigma}=2$.
The black shaded histogram shows the cumulative distribution of
distances $\Delta \theta_{\rm cl,peak}$ in arcmin for the
$\mathrm{\Lambda}$CDM light-cones for sources and clusters up to
$z=1.12$, while the green lines refers to the halo model predictions
when weak lensing maps are produced using our fast weak lensing model
(see Appendix).  The relative difference between those two histograms
remains well below $10\%$ and both distributions converge to unity
around $7$ arcmin. Nonetheless, peaks in the convergence maps created
using all the particles from the simulations have slightly misaligned
centres and less correlation with cluster centres than peaks in the
halo model maps because of the filamentary structure present in the
convergence field. This manifests also in the fact that peaks are not
spherical but typically elliptical.  The magenta histogram displays
the case of clusters with $z\leq 0.5$, that, as we will discuss later,
contribute the most to the convergence peaks with $f_{\sigma}=2$ and
$z_s=1.12$.  The vertical grey line indicates the angular scale of the
pixel of the convergence maps.  From the figure we notice that less
than five percent of the clusters have a centre of mass that overlaps
with the highest peak, approximately seventy percent are closer than
one arcmin to the highest peak while all clusters are within $7$
arcmin from some peak.  In order to see how the correlation between
clusters and peaks compares with respect to random points, in red we
display the cumulative distribution of the distance between clusters
and peaks, when the former are assumed to have random positions within
the field-of-view. The relative difference between the two
distributions clusters-peaks and random clusters-peaks (black and red
histograms, respectively) is displayed in the bottom panel. In this
panel we can notice in more details that at small scales clusters and
peaks are more correlated than random cluster positions which has a
maximum at about $15$ arcsec.

In Table~\ref{tab1} we summarise our results about the correspondence
of weak lensing peaks and clusters within the simulated past
light-cones.  Each row refers to a different cosmological model, while
the last one reports the findings in our halo model simulated fields
for the $\mathrm{\Lambda CDM}$ cosmology.  The numbers correspond to
the $25$ different light-cone realisations for each model for a total
of $625$ square degrees.

\section{Summary \& Conclusions}

In this work we have investigated the weak lensing peak statistics and
properties in a set of light-cones constructed from the coupled DM-DE
simulations of the $\mathrm{CoDECS}$ suite.  In particular we have
studied how the number density and area of weak lensing peaks differ
between models using typical source redshift distribution from ground
and space observations. In what follows we summarise our main
findings:
\begin{itemize}
  
\item the various cosmological models display different peak counts
  that increase with the signal-to-noise ratio $f_{\sigma}$. The
  extreme model $\mathrm{EXP003}$ for $f_{\sigma}=10$ displays a
  relative difference of about $20\%$ with respect to the
  $\mathrm{\Lambda CDM}$ and exhibits a different behaviour with
  respect to the $\mathrm{\Lambda CDM-HS8}$ which has the same power
  spectrum normalisation;
  
\item the fraction of area on the sky in peaks as a function of the
  signal-to-noise ratio displays a behaviour similar to that of the
  peak counts, except that for small values of $f_{\sigma}$ we found
  twin-peaks above a given threshold while for large values of
  $f_{\sigma}$ high convergence regions are isolated and become more
  distinct with respect to the projected linear and non-linear large
  scale matter density distribution; the relative difference between
  $\mathrm{EXP008e3}$ and $\mathrm{SUGRA003}$ in peak area is reversed
  with respect to peak counts underlining the importance of the
  concentration-mass relation in peak statistics;
  
  \item weak lensing peaks reflect the non-Gaussian properties of the
    underlying projected density field, trace non-linear structure
    formation processes and are very sensitive to the evolution of
    dark energy through the growth of density perturbations and the
    geometry of the expansion history.  This confirms the idea that
    weak lensing peak statistics, and their tomographic analysis, can
    provide complementary information to cosmic shear analysis alone;
    
\item peak abundance and properties are due to non-linear structures
  present along the line-of-sight and projected matter density
  distribution; in particular high signal-to-noise peaks are mainly
  produced by galaxy clusters and for the source redshift distribution
  as expected from a space-based experiment we find that almost the
  whole cluster population up to $z=0.5$ is in peaks with
  signal-to-noise ratio $f_{\sigma}=2$;

\item only five percent of the clusters have their centres of mass
  within the highest pixel in a peak of the convergence map
  (resolution $8.8$ arcsec).  On the other hand, all clusters are
  located within $7$ arcminutes of the maximum convergence pixel of a
  peak;
  
\item our halo model formalism for creating fast weak lensing
  simulations describes well the abundance of peaks for the different
  source redshift distributions;
  
\item the inclusion of substructures in our halo model raises the peak
  statistics only by a few percent;

\end{itemize}

Weak lensing peak statistics represents a powerful tool for
characterising non-Gaussian properties of the projected matter density
distribution. Peak properties depend on dark energy and their
tomographic analysis allows one to trace the structure formation
processes as a function of the cosmic time.  Our results underline the
necessity of combining peak statistics with other cosmological probes:
this will offer important results from upcoming wide field surveys and
will push cosmological studies toward new frontiers.
  
\label{summary}
\section*{Acknowledgments}                                           
CG and MB acknowledge support from the Italian Ministry for Education,
University and Research (MIUR) through the SIR individual grant
SIMCODE, project number RBSI14P4IH.  All authors also acknowledge the
support from the grant MIUR PRIN 2015 "Cosmology and Fundamental
Physics: illuminating the Dark Universe with Euclid". We acknowledge
financial contribution from the agreement ASI n.I/023/12/0 "Attivit\`a
relative alla fase B2/C per la missione Euclid".  MM and CG
acknowledge support from the Italian Ministry of Foreign Affairs and
International Cooperation, Directorate General for Country Promotion.
We thank also Federico Marulli and Alfonso Veropalumbo for useful
discussions.  We are also grateful to the anonymous reviewer for
her/his useful comments.

\appendix
	
\section{Fast Halo Model Simulations and a Model for Weak Lensing Peaks}

Modelling peak statistics represents a significant challenge when
using peak counts as complementary cosmological probe to cosmic shear
power spectrum. Predicting peaks in weak lensing convergence maps can
be done assuming that non-linear structures, like dark matter haloes,
are the main contributors to high-significance peaks.  In this paper
we have shown that while haloes hosting galaxy clusters are the main
contributors to high peaks, projection effects from small haloes
aligned along the line-of-sight contribute to peaks with low
signal-to-noise ratio.

In this appendix we will show that peaks identified in convergence
maps constructed using fast weak lensing simulations with
\textsc{WL-MOKA} \citep{giocoli17} are in very good agreement with
those in maps computed from full particle ray-tracing
simulations. Fast halo model simulations could prove extremely useful
by reducing the computational requirements for N-body simulations by
some orders of magnitude both in cosmic shear power spectrum and peak
statistics \citep{lin15a,lin15b,matilla16} when combined with
approximate simulation methods like COLA \citep{izard18} and PINOCCHIO
\citep{monaco13,munari17,monaco16}.  As discussed by \citet{giocoli17}
on a single light-cone simulation, our fast halo model method is
approximately $90$ per cent faster than a full ray-tracing simulation
using particles. However, it should be stressed that an N-body run of
$1$ Gpc/$h$ with $1024^3$ collisionless particles from $z = 99$ to the
present time using the GADGET2 code \citep{springel05a} takes around
$50$ $000$ CPU hours, while a run with an approximate method may take
approximately $750$ CPU hours to generate the past-light cone up to
the desired maximum redshift $z = 4$.\footnote{All the CPU times given
  here have been computed and tested in a $2.3$ GHz workstation.}

\begin{figure*}
  \includegraphics[width=0.95\hsize]{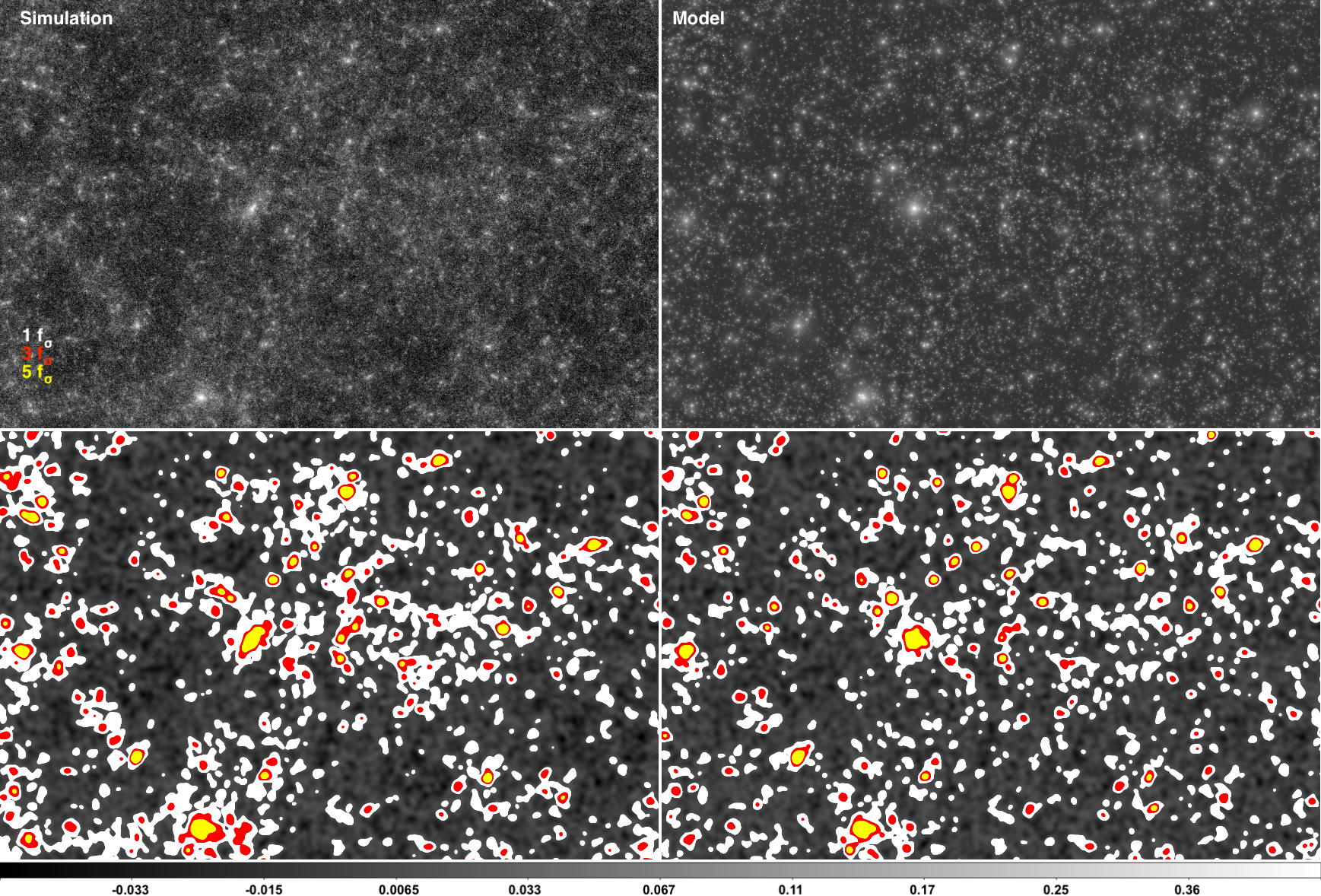}
  \caption{\label{figsimmodel}Top panels: convergence maps for source
    redshift $z_s=1.12$. In the left panel we show the map computed
    using all particles within the light-cone while on the right we
    display the reconstruction performed using all haloes with
    friends-of-friends mass larger than $2.1 \times 10^{12}
    M_{\odot}/h$. Bottom panels: peak detection in convergence maps
    created using particles (left) and haloes (right) from the same
    light-cone realisation of the $\mathrm{\Lambda CDM}$
    simulation. The convergence map has been constructed assuming
    $z_s=1.12$, noised and smoothed assuming $\sigma_F=1$ arcmin and
    $33$ galaxies per square arcmin.  In white, red and yellow we
    display the pixels in the map which are $1$, $3$ and $5$ times
    above the noise level.}
\end{figure*}

The theoretical approach for weak lensing peak prediction is based on
the projected halo model formalism \citep{cooray02}.  A full
characterisation of the halo population along the line-of-sight, with
consistent clustering properties, gives us the possibility of
predicting not only the peaks in cluster regions but also those in the
field , mainly due to projected interposed mass density distribution.
 
In order to build our peak model, in addition to the convergence maps
constructed using the particles from the numerical simulations, for
the $\mathrm{\Lambda CDM}$ model we also use a sample of maps computed
using the halo properties as presented in \citet{giocoli17}. In order
to do so, we use the corresponding projected halo and subhalo
catalogue from \textsc{MapSim}, considering all friends-of-friends
groups above the resolution $M>2.1\times 10^{12}M_{\odot}/h$. Each
halo, as read from the simulation catalogue and present within the
considered light-cone field-of-view, is assumed to be spherical and
characterised by a well defined density profile \citep{navarro96}
(hereafter NFW). We assume the halo concentration $c$ to be mass and
redshift dependent as in \citet{zhao09} in which we imply the mass
accretion history model by \citet{giocoli12b} and we assume a
log-normal scatter in concentration for fixed halo mass of
$\sigma_{\ln c}=0.25$ consistent with the results of different
numerical simulations \citep{jing00,dolag04,sheth04b,neto07}.  In this
case we can compute the convergence map by integrating the halo
profile along the line-of-sight up to the virial radius that can be
read as:
\begin{equation}
\kappa  (x,y)   =  \int_{-R_{vir}}^{R_{vir}}   \rho(x,y,z)  \mathrm{d}
z/\Sigma_{\rm crit}\,,\label{eqkappa}
\end{equation}
where
\begin{equation}
  \Sigma_{cr} \equiv \frac{c^2}{4 \pi G} \frac{1}{D_{lens}}
\end{equation}
is the critical surface mass density.  For the NFW profile and
assuming that along the line-of-sight we can integrate up to infinity,
equation~(\ref{eqkappa}) can be simplified to \citep{bartelmann96a}:
 \begin{equation}
 \kappa_{NFW}(x,y)=     \frac{2    \rho_s     r_s}{\zeta^2    -1     }
 \dfrac{F(\zeta)}{\Sigma_{\rm crit}}\,,
\end{equation}
where $\zeta\equiv\sqrt(x^2+y^2)/r_s$, $r_s=R_{vir}/c$, and:
\begin{displaymath}
  F(\zeta) = \left\{ \begin{array}{ll} 1- \frac{2}{\sqrt{\zeta^2-1}}
    \arctan\sqrt{\dfrac{\zeta-1}{\zeta+1}}&\zeta>1, \\ 1-
    \frac{2}{\sqrt{1-\zeta^2}}
    \mathrm{arctanh}\sqrt{-\dfrac{\zeta-1}{\zeta+1}} &\zeta<1, \\ 0&
    \zeta=1.\\
  \end{array} \right.
\end{displaymath}

Left and right panels of Fig.~\ref{figsimmodel} show the convergence
maps for $z_s=1.12$ of one light-cone realisation of the
$\mathrm{\Lambda CDM}$ model using particles and haloes,
respectively. The top panels show the convergence maps for $z_s=1.12$
while in the bottom we have includeed random noise assuming a number
density of galaxies $n_g=33$ arcmin$^{-2}$ and the maps have been
convolved with a Gaussian filter with $\sigma_F=1$ arcmin.  In white,
red and yellow we display the regions in the maps that are $1$, $3$
and $5$ times above the noise level. From the figure we notice that
qualitatively the peak location is very similar: the most massive
haloes are responsible for the highest convergence peaks, regions with
few systems appear, in projection, under-dense. However the shapes of
the peaks in the right panel are quite spherical as the haloes used in
the construction are, however the halo locations correlate with the
peaks as well as with the large scale matter density distribution
\citep{despali14,bonamigo15,despali17}.

\begin{figure}
  \includegraphics[width=1\hsize]{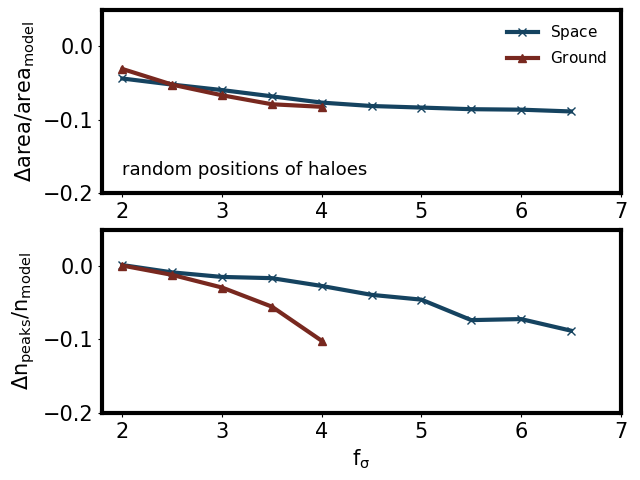}
  \caption{\label{figrandompos}Relative peak area and counts as a
    function of the noise threshold, from our fast lensing halo model,
    when the halo positions are read from the simulation and defined
    to be random.  The red and blue data points show the results
    considering a space or a ground based analysis, respectively.}
\end{figure}

In producing the lensing simulation model using haloes we have been
consistent in taking the halo positions from the simulation, and
projecting them on the plane of the sky.  This means that up to the
simulation scale of $1$ Gpc/$h$ the clustering of the systems is
preserved. However one may ask if this has a direct impact on the peak
counts of the constructed convergence maps.  In order to understand
this for each halo model light-cone we have created 16 realisations
where we have preserved the halo masses and concentrations but we have
assigned to each halo a random position within the field of view.  In
Figure~\ref{figrandompos} we display the relative number counts and
area in peaks as a function of the signal-to-noise level $f_{\sigma}$
between the halo model simulation when positions are read from the
simulation and when they are randomly assigned. We show the results
both for a space- and ground-based analysis displayed in blue and red,
respectively.  The figure shows that proper halo positions are
necessary for a good characterisation of the peak statistics mainly
for high values of the noise level. These allow a good description of
the large scale density distribution and of the effect of correlated
and uncorrelated structures on the location of high density
regions. From the figure we can see that for large values of the noise
level the relative difference in the area and in the number of peaks
tends to $10\%$ and that in the upper panel already for $f_{\sigma}=2$
the relative difference is about $5\%$.

\bibliographystyle{mn2e}

\bsp	
\bibliography{globalbibs}
\label{lastpage}
\end{document}